\documentstyle[12pt]{article}
\advance\textheight by 60pt
\advance\voffset by -40pt
\advance\textwidth by 50pt
\advance\oddsidemargin by -25pt
\advance\evensidemargin by -25pt

\def\single_space{\baselineskip 12pt plus 1pt minus 1pt}
\def\one_and_a_half_space{\baselineskip 19pt plus 1pt minus 1pt}
\def\double_spacesp{\baselineskip 24pt plus 2pt minus 2pt}

\newcommand{\geqnew}{\stackrel{>}{\!\ _{\sim}}}
\newcommand{\leqnew}{\stackrel{<}{\!\ _{\sim}}}

\begin{document}
\begin{titlepage}
\begin{flushright}
{\bf PSU/TH/231} \\
{\bf December 2000} \\
{\bf Revised, April 2001} \\
\end{flushright}
\vskip 1.5cm
\double_spacesp
{\Large
{\bf
\begin{center}
Expectation value analysis of  wave packet solutions 
for the quantum bouncer: short-term classical and long-term
revival behavior
\end{center}
}}
\vskip 1.0cm
\begin{center}
M.~A.~Doncheski\footnote{mad10@psu.edu} \\
Department of Physics \\
The Pennsylvania State University\\
Mont Alto, PA 17237  USA \\
\vskip 0.1cm
and \\
\vskip 0.1cm
R. W. Robinett\footnote{rick@phys.psu.edu} \\
Department of Physics\\
The Pennsylvania State University\\
University Park, PA 16802 USA \\
\end{center}
\vskip 1.0cm
\begin{abstract}
 We discuss the time development of Gaussian wave packet solutions
of the `quantum bouncer' (a quantum mechanical particle subject to
a uniform downward force,  above an impermeable flat surface). 
We focus on the evaluation and visualization of 
the expectation values and uncertainties of position and momentum
variables during a single quasi-classical period as well as during
the long term collapsed phase and several revivals. This approach
complements existing analytic and numerical analyses of this system, 
as well as being useful for comparison with similar results for the 
harmonic oscillator and infinite well cases. 
\end{abstract}
\end{titlepage}
\double_spacesp

\begin{flushleft}
 {\large {\bf I. Introduction}}
\end{flushleft}
\vskip 0.5cm

With the advent of modern computer technology, robust numerical 
calculations of time-dependent phenomena in quantum mechanics are now
common, as is the visualization of the resulting effects
\cite{thaller} -- \cite{styer_1}. The time-evolution of wave packet
solutions for many scattering geometries as well as for bound state
systems \cite{doncheski} are discussed with increasing frequency
 in the pedagogical literature,
illustrating not only such familiar aspects as wave packet spreading,
but also extending student experience to more
novel phenomena such as wave packet revivals. Reviews of wave packet
revivals (initially highly localized quantum wave packet solutions which 
exhibit quasi-classical behavior, then disperse or spread in time 
to a so-called collapsed phase, 
only to reform later to something very much like it's initial state)
in many familiar model one-dimensional quantum mechanical systems such
as the harmonic oscillator and infinite well have appeared
\cite{square_1} -- \cite{robinett_2} providing
students with accessible examples of an important quantum effect which
can be probed experimentally in atomic systems.

Besides studies of such effects in the two 'classic' quantum mechanical
model systems mentioned above, 
wave packet propagation, and especially the structure of 
revivals, has also recently been discussed in the quantum version of another
familiar classical system, the so-called quantum bouncer
\cite{bouncer_1}, \cite{bouncer_2}. This system is the bound state version of
the classical 'falling object', and is defined by the potential 
energy 
\begin{equation}
      V(z)  = \left\{ \begin{array}{ll}
               \infty & \mbox{for $z<0$} \\
               Fz & \mbox{for $0<z$}
                                \end{array}
\right.
\label{potential_definition}
\end{equation}
corresponding to a constant downwards force acting on a particle above
an impenetrable flat surface at $z=0$. Generalizing on many early
papers \cite{bouncer_3} which discuss the time-independent solutions for this 
problem, the author of Ref.~\cite{bouncer_1} focuses attention on deriving
expressions for the collapse and revival times for initially Gaussian
wave packet solutions, using a mixture of numerical and analytic techniques.
In this note, we will revisit this problem, along much the same lines as in
Ref.~\cite{robinett_2}, by examining, in detail, the short-term
(quasi-classical) and long-term (revival) time-development of such
solutions in terms of their position and momentum expectation values and
uncertainties. This type of expectation value analysis, coupled with 
existing analytic, numerical, and visualization studies can then help 
form a more complete picture of the highly non-trivial time-development 
possible in one of the 'classic' one-dimensional model systems in 
introductory quantum mechanics. 

As often happens, the more familiar and tractable cases of the 
harmonic oscillator and the infinite well examples exhibit special features, 
in this case, in terms of their time-dependence, compared to a more general 
form such as in Eqn.~\ref{potential_definition}. For that reason, in
Sec.~II, we first very briefly review Gaussian wave packet solutions 
in these two  more familiar systems. The bouncer system shares with the
infinite well the feature of the 'bounce' at the impenetrable wall, but
some observables, such as the time-dependent wave packet spread, 
exhibit a much more typical cyclic structure than for the infinite well, 
where only the free-particle  spreading time \cite{robinett_2} is of 
relevance. Since the numerical studies presented here
can also make contact with 
analytic solutions to the problem of a particle undergoing uniform 
acceleration, we also mention those solutions in Sec.~III and then
proceed to  detailed  results for the quantum bouncer.

\begin{flushleft}
 {\large {\bf II. Gaussian wave packet solutions for the
harmonic oscillator and infinite well}}
\end{flushleft}
\vskip 0.5cm

Because of the special nature of the energy level structure of the
harmonic oscillator, any time-dependent solution of this problem,
$\psi(x,t)$,  will
be exactly periodic and return precisely to its initial state after
the classical period, $T_{cl} = 2\pi/\omega$. Using standard
propagator techniques or other methods, it is, in fact, easy to construct
closed form Gaussian wave packet solutions \cite{saxon} such as
\begin{eqnarray}
|\psi(x,t)|^2 &= &\frac{1}{\sqrt{\pi}L(t)}
e^{-[x-x_0\cos(\omega t)]^2/L^2(t)} 
\label{other_sho_solutions}
\\
|\phi(p,t)|^2 &  =& \frac{1}{\sqrt{\pi}p_L(t)}
e^{-[p+m\omega x_0\sin(\omega t)]^2/p_L^2(t)}
\label{sho_wavepacket_solutions}
\end{eqnarray}
which have expectation values which satisfy the classical equations of
motion, namely
\begin{equation}
\langle x \rangle_t  =  x_0 \cos(\omega t)
\qquad
\mbox{and}
\qquad 
\langle p \rangle_t = -m\omega x_0 \sin(\omega t).
\end{equation}
The fact that the expectation values for any wave packet solutions for
the harmonic oscillator  satisfy the classical equations of motion
does not depend on these specific Gaussian forms, but can be shown to be
true analytically in a quite general way \cite{styer_2}.
The solutions in Eqns.~(\ref{other_sho_solutions}) and 
(\ref{sho_wavepacket_solutions}) have time-dependent
position and momentum uncertainties given by 
$\Delta x_t = L(t)/\sqrt{2}$ and $\Delta p_t = p_L(t)/\sqrt{2}$ where
\begin{eqnarray}
[L(t)]^2   & = & L^2 \cos^2(\omega t) + \left(\frac{\hbar}{m\omega L}\right)^2 
\sin^2(\omega t)  \\
\label{sho_position_spread}
[p_L(t)]^2 & = & \left(\frac{\hbar}{L}\right)^2 \cos^2(\omega t) + 
(Lm\omega)^2 \sin^2(\omega t)
\end{eqnarray}
and the parameter $L$ sets the scale for both the position- and 
momentum-space spreads. Unless $L = \sqrt{\hbar/m\omega}$, 
these position- and momentum-uncertainties  oscillate with a 
period twice that of the classical motion \cite{saxon}, 
regaining the initial values of $\Delta x_0$ and $\Delta p_0$ at two 
opposing points in phase space. 
Once again, these results are not specific to the Gaussian solution,
but have been derived quite generally in Ref.~\cite{styer_2}:
they are also sometimes rediscovered \cite{pulsating_1}, \cite{pulsating_2} 
in these specific cases. 


Another tool which is standardly used to probe the wave packet's 
approximate return to the initial 
state is the auto-correlation function \cite{nauenberg}, defined by
\begin{equation}
A(t)  =  \int_{-\infty}^{+\infty} \psi^*(z,t) \,\psi(z,0)\,dz 
 = \int_{-\infty}^{+\infty} \phi^*(p,t)\, \phi(p,0)\,dp 
 =  \sum_{n=1}^{\infty} |c_n|^2 \,e^{iE_nt/\hbar} 
\label{autocorrelation_function}
\end{equation}
which measures the overlap of the position- or momentum-space wavefunction
at later times with the initial state. For the oscillator, one can argue
that because the wave packets are exactly periodic and never collapse,
there are no revivals.

On the other hand, 
for the infinite well, because the energy levels are integral multiples of
a common value (but not equally spaced), 
there are exact revivals \cite{square_1}, but initial
Gaussian-type wave packets do undergo dispersion into a collapsed phase. 
For example, wave packets constructed from energy eigenstates via 
\begin{equation}
\psi(x,t) = \sum_{n=0}^{\infty} c_n u_n(x) e^{-iE_nt/\hbar}
\label{standard_expansion}
\end{equation}
where $u_n(x) = \sin(n\pi x/L)/\sqrt{L}$ for the infinite well, 
with quasi-Gaussian expansion coefficients of the form
\begin{equation}
c_n = 
\sqrt{
\frac{\alpha \hbar \sqrt{\pi}}{L}}
e^{-\alpha^2 (p_n-p_0)^2} e^{ip_nx_0/\hbar}
\end{equation}
(where $p_n = n\hbar \pi /L$)
give an initial momentum distribution very close to the
standard one for Gaussian free-particle packets, namely
\begin{equation}
\phi_0(p) = \frac{\alpha}{\sqrt{\pi}}e^{-\alpha^2(p-p_0)^2/2} 
e^{ipx_0/\hbar}
\end{equation}
and yield  localized Gaussian-like position wave packets which are
initially very close to the free-particle form
\begin{equation}
|\psi(x,t)|^2 = \frac{1}{\beta_t \sqrt{\pi}}
e^{-(x-[x_0 + p_0t/m])^2/\beta_t^2}
\end{equation}
where $\beta _t = \hbar \alpha \sqrt{1+t^2/t_0^2}$ with 
$t_0 \equiv m \hbar \alpha^2$. Such packets undergo quasi-classical
motion, bouncing back and forth between the two infinite walls, with
an increasing width given roughly by an envelope defined by the
time-dependent free-particle width, 
$\Delta x_t = \beta_t/\sqrt{2}$ \cite{robinett_2}. During the impulsive 
collisions with the walls, 
the position-space width
temporarily decreases during the time that the momentum-space distribution
is `flipping' from positive to negative values. 

As mentioned above, 
since the quantum bouncer has features in common with both systems, namely
one infinite wall boundary at which 'bounces' will occur and a second,
smoother potential barrier, we will keep both these cases
in mind as we examine  the short- and long-term behavior of expectation
values for the quantum bouncer.

\begin{flushleft}
 {\large {\bf III. Expectation values for Gaussian wave packets
for the quantum bouncer}}
\end{flushleft}
\vskip 0.5cm

Before turning to wave packet solutions of the quantum bouncer
problem, we first recall some results from the related problem of a
quantum particle undergoing uniform acceleration, namely subject to
a constant force, $F$, with a potential given by $V(x) = -Fx$ everywhere
in space. Transforming the resulting Schr\"odinger equation into 
momentum-space, one can construct  arbitrary solutions \cite{robinett_book} 
of the form
\begin{equation}
\phi(p,t) = \phi_0(p-Ft) \, e^{i[(p-Ft)^3 - p^3]/6mF\hbar}
\end{equation}
where $\phi_0(p)$ is any initial momentum distribution. This result 
already implies
that the momentum distribution for this case simply translates uniformly
in time, with no change in shape,  since
\begin{equation}
|\phi(p,t)|^2 = |\phi_0(p-Ft)|^2
\label{momentum_translation}
\end{equation} 
so that $\Delta p_t = \Delta p_0$. 
(We expect to see this behavior initially in the quantum bouncer case,
until the wave packet nears the wall,  as  well as after the 
collision, at least for the first few classical periods.)
 Using this momentum-space solution, we can also construct closed-form
position-space solutions; for example using the distribution
\begin{equation}
\phi_0(p) = \sqrt{\frac{\alpha}{\sqrt{\pi}}}
e^{-\alpha^2(p-p_0)^2/2} e^{ipx_o/\hbar}
\end{equation}
we find  a general Gaussian wave packet for the accelerating particle
\begin{equation}
|\psi(x,t)|^2 = \frac{1}{\beta_t \sqrt{\pi}}
e^{-(x-[x_0 + p_0t/m+Ft^2/2m])^2/\beta_t^2}
\label{uniform_acceleration_packet}
\end{equation}
with $\beta_t = \alpha \hbar \sqrt{1+t^2/t_0^2}$. (The same results
can be obtained in a variety of ways, using operator 
\cite{robinett_acceleration} or propagator 
\cite{holstein_acc} methods or other techniques \cite{new_acc}.)
This solution gives 
\begin{equation}
\langle x \rangle_t  =   x_0 + \frac{p_0 t}{m} + \frac{F}{2m} t^2
\qquad
\quad
\mbox{and}
\qquad
\quad 
\Delta x_t           =  \frac{1}{\sqrt{2}} \beta_t = 
\frac{\alpha \hbar}{\sqrt{2}} \sqrt{1+t^2/t_0^2}.
\label{accelerating_packet}
\end{equation}
Thus, the expectation value tracks the classical trajectory, while the
wave packet spreads in exactly the same way as the standard free-particle
Gaussian and we expect to see remnants of this behavior as well.

The author of Ref.~\cite{bouncer_1} has constructed quasi-Gaussian
wave packet solutions for the quantum bouncer 
using a combination of numerical and semi-analytic techniques,
focusing on the derivation of quantities such as 
approximate revival time, $T_{rev}$. 
(See Ref.~\cite{square_1} for a very general derivation of both the classical
period and quantum revival time for arbitrary bound state energy spectra). 
For an initial wave packet which is constructed so as 
to represent a particle released from rest at a height $z_0$, the
classical period of motion, $T_{cl}$,  and the revival time,
$T_{rev}$,  are given by
\begin{equation}
T_{cl} = 2\sqrt{\frac{2mz_0}{F}}
\qquad
\mbox{and}
\qquad
T_{rev} = \frac{4}{\pi} \left(\frac{2mz_0^2}{\hbar}\right)
\label{revival_time}
\end{equation}
respectively. 
For simplicity, following Ref.~\cite{bouncer_1}, we take units such
that 
\begin{equation}
\hbar = 1,
\qquad
\quad
2m = 1,
\qquad
\mbox{and}
\qquad 
F = 1
\end{equation}
and we choose particular initial values, namely
\begin{equation}
z_0 = 25
\qquad
\quad
\mbox{and}
\quad
\qquad 
\Delta z_0 = 1.
\label{wave_packet_parameters}
\end{equation}
With these values, the classical period and revival time are given by
\begin{equation}
T_{cl} = 10
\qquad
\mbox{and}
\qquad 
T_{rev} = \frac{4}{\pi}(5)^2 = 795.78.
\end{equation}
The collapse time, the time it takes for the initially localized
wave packet to de-phase and become spread over the entire well, is given,
in these dimensionless units as well as for our specific wavepacket 
parameters, as
\begin{equation}
T_{coll} = \frac{T_{cl}^3}{8\Delta z_0}
\qquad 
\quad
\mbox{or}
\qquad
\quad 
T_{coll} = 125
\label{collapse_time}
\end{equation}
respectively.

Using these parameters, we can evaluate both the position-space and 
momentum-space wavefunctions in a manner which is similar to that in 
Ref.~\cite{bouncer_1}. We generate the expansion coefficients, $c_n$, 
in Eqn.~(\ref{standard_expansion}) by explicit numerical calculation
of the overlap integrals of the initial Gaussian form with the normalized 
energy eigenstates obtained by numerical solution of the Schr\"odinger 
equation; the eigenstates, normalization constants, and expansion
coefficients are then compared to the  Airy function 
solutions used in Ref.~\cite{bouncer_1}, supplemented by the more exact 
analytic results in Ref.~\cite{bouncer_2}, as a cross-check.
(We truncate the expansion in eigenstates at a level consistent with
the double-precision accuracy of our numerical evaluation, typically
at values of the $c_n \approx 10^{-6}$; this is to be be compared to the
maximum values of $c_n \sim 0.5$. The resulting wavefunctions are then
found to be appropriately normalized to unity to within $10^{-6}$ or better 
over all time intervals considered here.)

We use this more numerical procedure partly because this  technique runs 
more efficiently in our computing
environment, but also because we intend to eventually extend our investigations
of wave packet behavior to other, more general power law potentials of the 
form $V_{(k)}(x) = V_0|x/a|^k$ for comparison to recent work on the
revival times in such 
potentials \cite{robinett_jmp}. This class of potentials
contains the harmonic oscillator ($k=2$) and the infinite well ($k = \infty$)
as special cases and can also be analyzed with an infinite wall added at
$x=0$ to give  the quantum bouncer case (where $k=1$): this fact
also helps motivate  our brief review of those two special cases.

We then plot the corresponding probability distributions over
the first classical period in Fig.~1. On the left, the position-space
probability is  seen to 
spread in a manner which is numerically consistent with 
Eqn.~(\ref{uniform_acceleration_packet}), while the calculated position
value $\langle z \rangle_t$ (solid curve) agrees well with the classical
expectation for the trajectory (dashed) except, of course, for the 
cusp at the 'bounce'.  The packet exhibits the standard `interference'
pattern during the collision with the wall \cite{doncheski},
\cite{andrews}, at the 'bounce', and then reforms into something like the
initial packet (compare to the dotted initial packet superimposed
on the $t=10$ case),  only wider.

For the momentum-space distributions (shown on the right of Fig.~1), we
also see features of both the classical motion and the uniformly accelerated
wave packet. The expectation value of momentum $\langle p \rangle_t$,
calculated from $|\phi(p,t)|^2$ and plotted as the solid curve, is 
once again consistent with the classical trajectory (dashed curve), 
except near the
discontinuous, impulsive change in momentum values at the `bounce'. The
shape of the momentum-distribution follows the form expected from
Eqn.~(\ref{momentum_translation}), namely uniform translation with no
change in shape, from $t=1 \rightarrow  t=3$ and then again from
$t=8 \rightarrow  t = 10$, that is, during the time when it is not in 
collision with the wall, but with a definite final change in shape, 
compared to the initial $|\phi_0(p)|^2$ superimposed on the $t=10$ result. 
The dotted vertical lines indicate the values of $p=0$, and  also the 
classically expected minimum and maximum values of momentum given by
$\pm p_M = \pm \sqrt{2mE} = \pm \sqrt{2mFz_0}$. 

The asymmetric shape of $|\phi(p,t)|^2$ which is obvious after one period 
can be understood using purely classical arguments. For example, we can
examine the classical trajectories of particles released from 
the same height, $H$, but with varying initial velocities (or momenta);
the resulting time-dependent momenta, $p(t)$ versus $t$, are shown in Fig.~2.
We can see the same asymmetry arising at $t= T_{cl}$ as for the
quantum mechanical $|\phi(p,t)|^2$ in Fig.~1; trajectories with momentum 
values  $p \leqnew 0$ at $t = 0$ are shifted upwards towards $p \approx 0$ 
at $t = T_{cl}$,  while those with $p \geqnew 0$ are also moved to higher 
values to form the observed high-momentum 'tail'. (A very similar classical 
explanation of the time-dependent behavior of the momentum-space probability 
density for an otherwise free particle undergoing a 'bounce' from an infinite 
wall is discussed in Ref.~\cite{doncheski}; in that case, the observed
asymmetry in $|\phi(p,t)|^2$ during the 'bounce' is due to the fact that
the fastest (highest momentum) components of the wavepacket strike the 
wall first, and hence are reversed in direction, before the slow components.)

We next focus on the time-development of quantum mechanical expectation 
values over
the first eight classical periods and plot $\langle z \rangle_t$ and
$\langle p \rangle_t$ over this interval in Fig.~3. 
Once again, the solid curves
are the quantum mechanical expectation values, while the dashed curves
are the classical trajectories. Clearly, the  quantum results generally
track the classical values, but become increasingly smoothed  out
as the packet spreads in time. We show in the same figure the time-development
of the autocorrelation function, $|A(t)|^2$ versus $t$, over the same
time interval which shows the (increasingly) approximate nature of the 
return to the initial state at multiples of the classical period: these 
results look very similar to data for wave packets in the infinite well 
\cite{square_1}, \cite{robinett_2} as it approaches the  collapsed state, 
while the same plot for the oscillator gives a completely periodic structure
with $A(t)$ returning exactly to $1$ at integral multiples of $\tau =
2\pi/\omega$.

In Fig.~4, we plot the time-dependent uncertainties in the position 
($\Delta z_t$, top) and momentum ($\Delta p_t$, bottom) over the first
eight classical periods. We note that the behavior of $\Delta z_t$ is
somewhat similar to that for the oscillator case: it initially increases in 
a manner consistent with Eqn.~(\ref{accelerating_packet}) for much of the
first half period, but then shrinks dramatically near $T_{cl}/2$ during the
collision with the wall (as in Ref.~\cite{doncheski}), 
and finally shrinks more slowly during the second
half period, returning to something close to its initial spread. This
behavior is  different than the case of the infinite well 
where the wavepacket spreading is mostly determined by the free-particle
expansion and more similar to the harmonic oscillator case 
in Eqn.~(\ref{sho_position_spread}) and general bound state systems.

In Fig.~4,
the expectation value during each period is superimposed with two
expressions for the spread (shown as pairs of dashed curves);
specifically, for the first period we use
\begin{equation}
\Delta z_t = \Delta z_0 \sqrt{1 + \left(\frac{t}{t_0}\right)^2}
\qquad
\mbox{and}
\qquad
\Delta z_t = \Delta z_0 \sqrt{1 + \left(\frac{T_{cl}-t}{t_0}\right)^2}
\label{first_cycle_spreads}
\end{equation}
and we extend these  cyclically to later periods as shown in the figure.
These agree fairly well with the observed time-dependence during the first 
cycle except during the collision phase, but become increasing bad 
representations as the wave packet spreads into its collapsed phase. Thus,
over at least the first few periods, the return to the initial spread is
similar to that in Eqn.~(\ref{sho_position_spread}). 

In the same figure, the spread in momentum is also illustrative of both
classical motion and results for accelerated wave packets. Over the first 
cycle, when the packet is not colliding with the wall, the spread in momentum
remains constant, as expected from Eqn.~(\ref{momentum_translation}), while it
increases dramatically during the collision time, just as with the
`bouncing wavepacket' in Ref.~\cite{doncheski}. After each bounce, however,
the spread in momentum has grown  slightly so that $\Delta p_t$ increases
in a quasi-stepwise manner between collision, with flat plateaus between
each impulsive `spike'.

We now turn our attention to the long-term behavior of these solutions
and plot in Fig.~5 the same quantities, $\langle z \rangle_t$,
$\langle p \rangle_t$, and $|A(t)|^2$, as in Fig.~3, but now over an interval
containing two expected 
revivals; these are indicated by the bold vertical dashed lines.
We also indicate the collapse time, given by Eqn.~(\ref{collapse_time}),
after $t=0$ as well as one unit of collapse time around each expected
revival. The
reformation of the wavepacket at the revivals in clearly apparent in 
all three variables, as is the approach to the more constant values during
the collapsed phases. 

The horizontal dashed lines around which the quantum mechanical expectation
values cluster during the collapsed phases are given by the average values
derived from purely classical probability densities. Using standard
arguments about how much time a particle spends in a small interval of
position, we can find the normalized position-space classical probability
as
\begin{equation}
P_{CL}(z) = \frac{1}{2\sqrt{A(A-z)}} 
\qquad
\qquad
\mbox{for $0<z<A$}
\label{classical_position_probability}
\end{equation}
and which vanishes for all other values of $z$. The classical turning
point, $A$,  is given by $E = FA$, and  we naturally associate $A$ 
with the initial position $z_0$ in our quantum analysis. With
this identification, the classical average values are
\begin{eqnarray}
\langle z \rangle & = & \int_{0}^{z_0} z \, P_{CL}(z)\,dz = \frac{2}{3}z_0
 \nonumber \\
\langle z^2 \rangle & = &  \int_{0}^{z_0} z^2 \, P_{CL}(z)\,dz = 
\frac{8}{15}z_0^2
\label{classical_position_averages} \\
\Delta z & = &  \sqrt{\langle z^2 \rangle - \langle z \rangle^2} = 
\frac{2}{\sqrt{45}} z_0. \nonumber 
\end{eqnarray}
For the momentum-space probability distribution, we note that for a
constant force, the probability density is uniform over the allowed space
of values, namely
\begin{equation}
P_{CL}(p)  =  \frac{1}{2p_M}  
\qquad
\quad
\mbox{for}
\quad
\quad
-p_M < p < +p_M
\end{equation}
and vanishing for $|p|> p_M$: in  this expression we have
$p_M = \sqrt{2mE} = \sqrt{2mFA} = \sqrt{2mFz_0}$. 
The classical averages are then given by 
\begin{eqnarray}
\langle p \rangle & = &  0  \nonumber \\ 
\langle p^2 \rangle & = &  \frac{1}{3}p_M^2  
\label{classical_momentum_averages}\\ 
\Delta p & = &  \sqrt{ \langle p^2 \rangle - \langle p \rangle^2}
= \frac{1}{\sqrt{3}} p_M. \nonumber 
\end{eqnarray}
These results are known to agree well (after local averaging) with
the position- and momentum-space probability densities corresponding to
time-independent energy eigenstates for this type of potential 
\cite{robinett_probability}. We would expect that the quantum wave packet
expectation values to closely match these classical results (again, after
local averaging) during the collapsed phase as well, since the wave function is
closer to being an incoherent sum of many such stationary states 
rather than the
highly coherent, well-localized, quasi-periodic state close to the
initial time and later revivals. The time-dependent values of
$\langle z \rangle_t$ and $\langle p \rangle_t$ in Fig.~5 do 
collapse to just these values.

For the autocorrelation function  we have no classical analog, 
but we can formally evaluate $|A(t)|^2$ in terms of the expansion
coefficients, $c_n$,  of the wave packet. In the limit that the various
time-dependent  components of the packet get out of phase, we have the
expectation that  during the collapsed interval
\begin{equation}
|A(t)|^2   =   \left|\sum_{n=1}^{\infty} |c_n|^2 \,e^{iE_nt/\hbar}\right|^2 
 =  \sum_{n,m=1} |c_n|^2 |c_m|^2 \, e^{i(E_n-E_m)t/\hbar}
 \longrightarrow  \sum_{n=1} |c_n|^4 
\label{autocorrelation_average_value}
\end{equation}
as the `off-diagonal' terms tend to cancel as they have lose the
 phase coherence built into the initial state. This limiting value,
calculated using the numerically determined  $c_n$,  is plotted
in Fig.~5 (as the bold horizontal dashed line) and compares well to the
locally averaged value of $|A(t)|^2$ during much of the collapsed period.

In Fig.~6, we plot the position- and momentum-spreads, as in Fig.~4, but
once again over the longer time interval, and superimpose the classical
(constant) expectation values for $\Delta z$ and $\Delta p$ as
bold horizontal dashed lines.  In both plots, the quantum wave packet
expectation values and uncertainties cluster appropriately around the
classical results of Eqns.~(\ref{classical_position_averages})
and (\ref{classical_momentum_averages}) 
during the collapsed phase as expected.

Finally, one of the most interesting results of Ref.~\cite{bouncer_1} is
that the wave packet revivals happen in such a way that the probability
densities are almost completely out of phase with the classical motion. In
order to visualize this effect in the expectation value 
approach followed here, we plot
again the same quantities as in Fig.~3, namely $\langle z \rangle_t$,
$\langle p \rangle_t$, and $|A(t)|^2$, also over a time interval
corresponding to 8 classical periods, but now centered around the expected
revival time. In each case we indeed see that the quantum expectation 
values (solid) are approximately half a cycle out of phase with the 
corresponding classical trajectory values (dashed) and that the
autocorrelation function is peaked at regular intervals, but offset
from integral multiples of the classical period by roughly
$T_{cl}/2$. 

In conclusion, we have reexamined the interesting problem of quasi-Gaussian
wave packets for the quantum bouncer, focusing on the numerical calculation 
and visualization of the results through the expectation values
and uncertainties of position and momentum. We have focused on a single
classical period to illustrate the important similarities and significant
differences  between
this case and well-studied  periodic wave packet solutions to other
familiar problems such as the harmonic oscillator and infinite well as well
as to the closed-form solutions for the constant acceleration case in
Eqns.~(\ref{uniform_acceleration_packet}) and (\ref{accelerating_packet}). We
have also examined these quantities over the first few periods, well
before the collapsed phase and compared them to classical trajectory 
expectations.
Finally, we have confirmed, numerically and through visualization,
that the quantum expectation values approach
(in a locally averaged sense) the purely classical results from the
time-independent probability distributions in 
Eqns.~(\ref{classical_position_averages})
and (\ref{classical_momentum_averages}) during the collapsed phase as well
as illustrating 
the expected phase relationships between the classical trajectories and quantum
expectation values near the revivals.

\begin{flushleft}
 {\large {\bf Acknowledgments}}
\end{flushleft}
\vskip 0.5cm

The work of MAD was supported, in part, by the Commonwealth College of The
Pennsylvania State University under a Research Development Grant (RDG); 
the work of RR was supported, in part, by NSF grant DUE-9950702.

\newpage
\begin{flushleft}
{\Large {\bf 
Figure Captions}}
\end{flushleft}
\vskip 0.5cm
 
\begin{itemize}
\item[Fig.\thinspace 1.] Gaussian wave packet solutions for the
quantum bouncer in position-space ($|\psi(z,t)|^2$ versus $z$, left)
and momentum-space ($|\phi(p,t)|^2$ versus $p$, right) for various times
during the first classical period. The solid curves represent the
time-dependent expectation values of position  ($\langle z \rangle_t$, left) 
and momentum ($\langle p \rangle_t$, right) for these solutions.
The similar dashed curves are the 
classical trajectories, $z(t)$ (left) and $p(t)$ (right),  superimposed.
The wave packet parameters in Eqn.~(\ref{wave_packet_parameters}) are used.
For the momentum-space figure, the vertical dotted lines represent the values
$p=0$ and the classical extremal   values of momentum,  $\pm p_M = \pm 
\sqrt{2mFz_0}$.
\item[Fig.\thinspace 2.] Plots of the classical momentum versus time,
$p(t)$ versus $t$, for objects released from a height $H$ with initial
momenta given by $p(0) = 0$ (solid), $p(0) = \pm \Delta p$ (dotted), and 
$p(0) = \pm 2\Delta p$ (dashed). After one classical period, the distribution 
of momentum values is asymmetric about $p=0$, in just the same way as
the $|\phi(p,t)|^2$ plot in Fig.~1; compare the momentum values defined by 
the $t=0$, $(+2\Delta p, -2\Delta p)$ 'band' to the corresponding values 
at $t = T_{cl}$.
\item[Fig.\thinspace 3.] The solid curves represent the expectation
values of position ($\langle z \rangle_t$, top) and momentum
($\langle p \rangle_t$, middle) and the autocorrelation function
($|A(t)|^2$, bottom) versus time for the first eight  classical periods
of motion for the bouncing wave packet. The wave packet parameters in 
Eqn.~(\ref{wave_packet_parameters}) are used. The dashed lines are the
classical trajectories for comparison. The horizontal line in the top
figure corresponds to the average value of position over one classical
period, namely $\langle z \rangle = 2z_0/3$,  from 
Eqn.~(\ref{classical_position_averages}). 
\item[Fig.\thinspace 4.] Same as Fig.~3 except that the time-dependent
values of the uncertainty in position ($\Delta z_t$, top) and momentum
($\Delta p_t$, bottom) are shown over the first eight  periods. The dashed
curves in the top picture correspond to  the spreading (anti-spreading)
of a Gaussian wave packet as it 'slides' downhill (uphill) as in
Eqn.~(\ref{first_cycle_spreads}); these expressions
agree well with the observed values of $\Delta z_t$ during the first period,
except during those times when the packet is `bouncing' against the
infinite wall at $z=0$, near $T_{cl}/2 = 5$.  The dashed line in the 
$\Delta p_t$ figure (bottom) indicates the classical spread in momentum 
defined in Eqn.~(\ref{classical_momentum_averages}), $\Delta p = p_M/\sqrt{3}$.
\item[Fig.\thinspace 5.] Same as Fig.~3, but now over a longer time scale,
containing two revivals. The expected locations of the revivals, according
to Eqn.~(\ref{revival_time}), are shown as vertical dashed lines, and times
within one collapse time, $T_{coll}$, of the initial time and these 
two revival times
are also shown as vertical dotted lines. The classical average values
of position and momentum ($\langle z \rangle = 2z_0/3$ and $\langle p \rangle
= 0$) from Eqns.~(\ref{classical_position_averages}) and
(\ref{classical_momentum_averages}) are shown as bold horizontal dashed lines, 
as is the expected value of
the autocorrelation function during the collapsed phase, 
$|A(t)|^2 = \sum_{n} |c_n|^4$, from 
Eqn.~(\ref{autocorrelation_average_value}), using the numerically determined
values of the $c_n$. 
\item[Fig.\thinspace 6.] Same as Fig.~3, expect for the long term values of
$\Delta z_t$ and $\Delta p_t$. The classical average values of these
quantities
($\Delta z = 2z_0/\sqrt{45}$ and $\Delta p = p_M/\sqrt{3}$) 
 from Eqns.~(\ref{classical_position_averages}) and
(\ref{classical_momentum_averages}) are shown as bold horizontal dashed lines.
\item[Fig.\thinspace 7.] Same as Fig.~3, except for an eight classical period
time window around the first revival which is indicated by the
bold vertical dashed line.  Note how the quantum mechanical 
averages, $\langle z \rangle_t$ and $\langle p \rangle_t$ (solid curves),
 are out of phase with the classical trajectory values (dashed curves). In
the bottom figure, the dashed curve corresponds to the time-development of
the auto-correlation function, $|A(t)|^2$, over the first eight periods
of the motion, for comparison. 
\end{itemize}

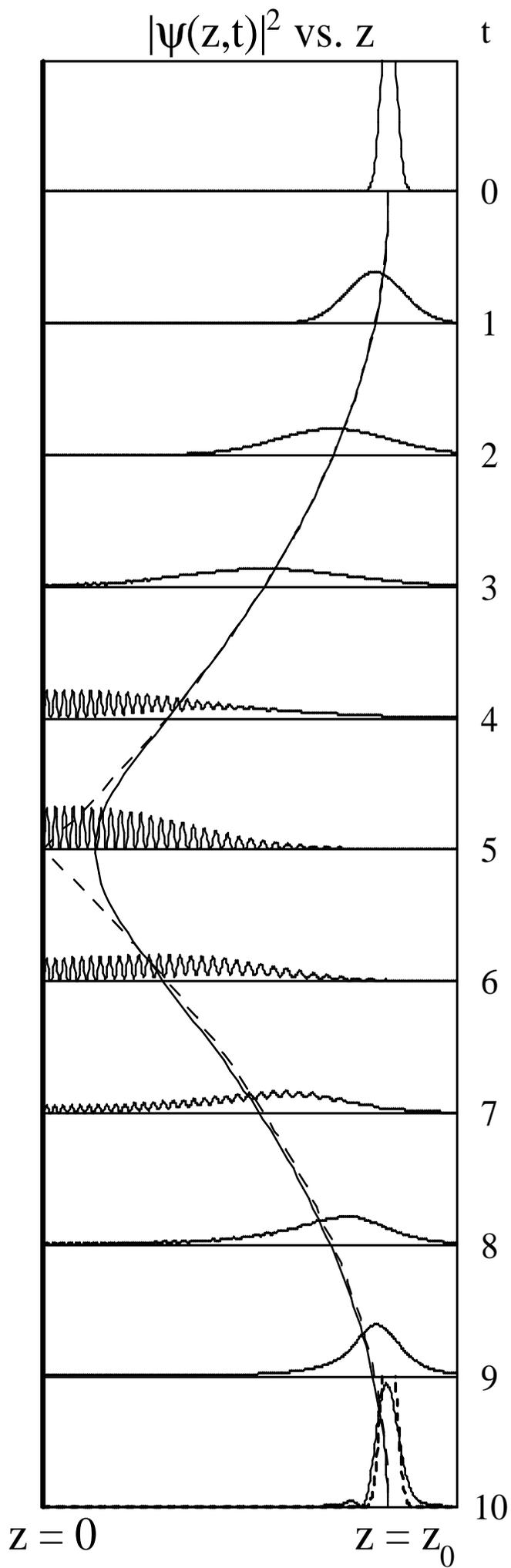
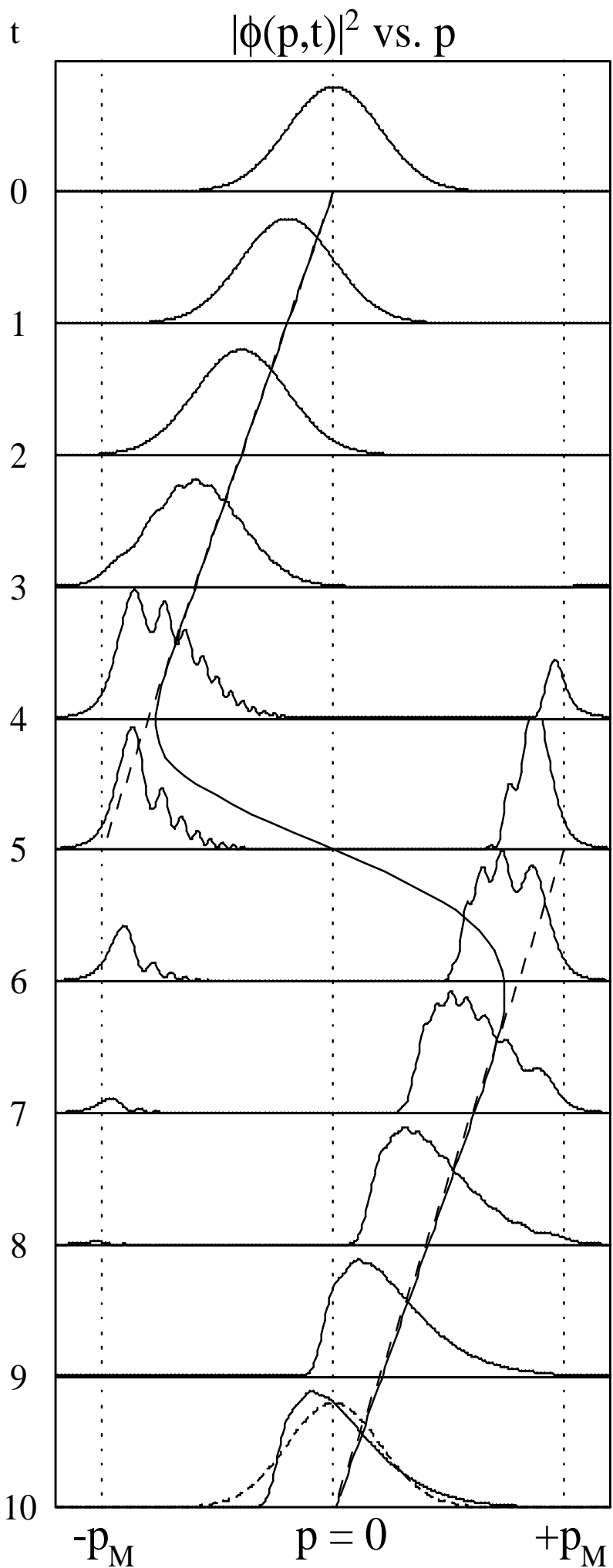
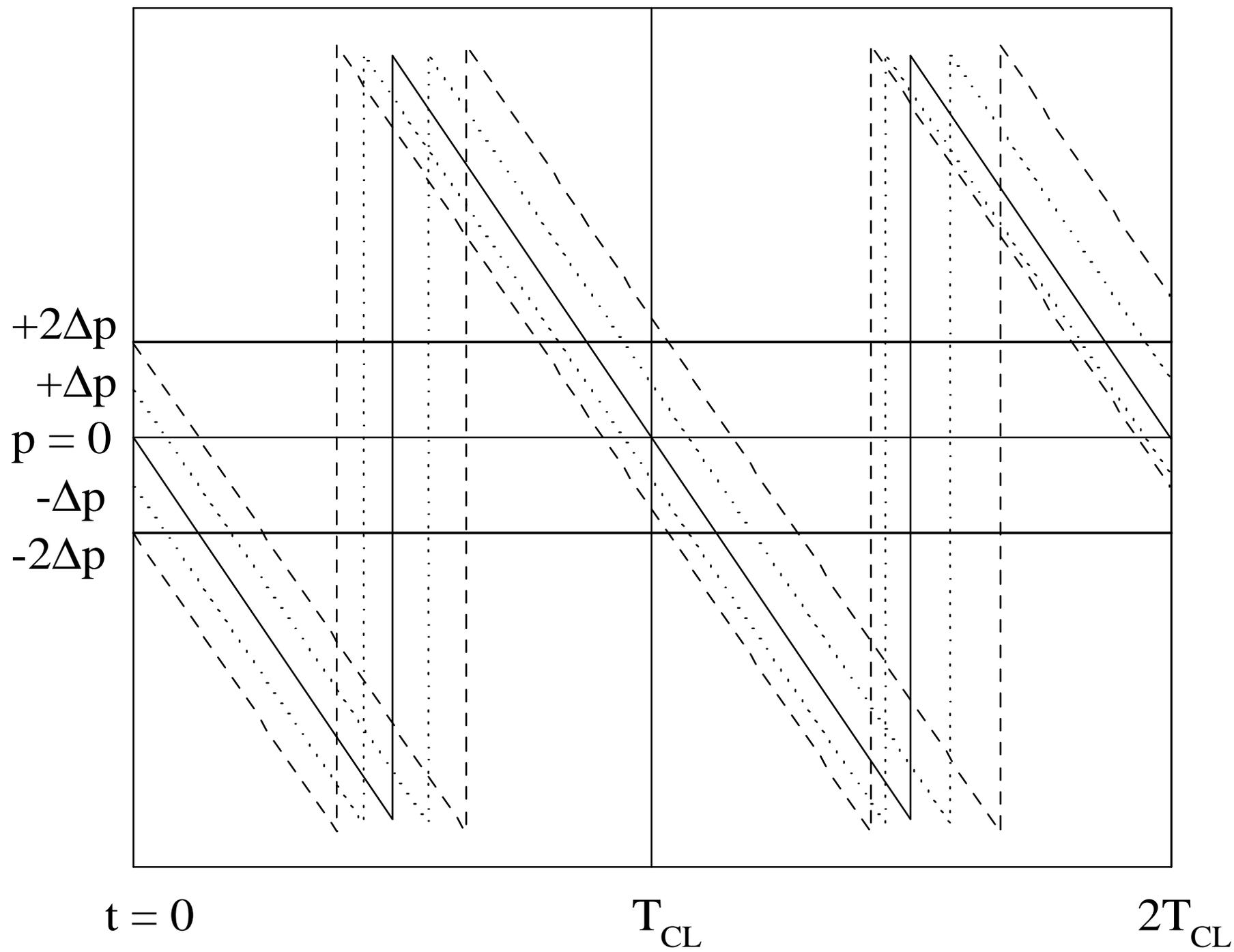
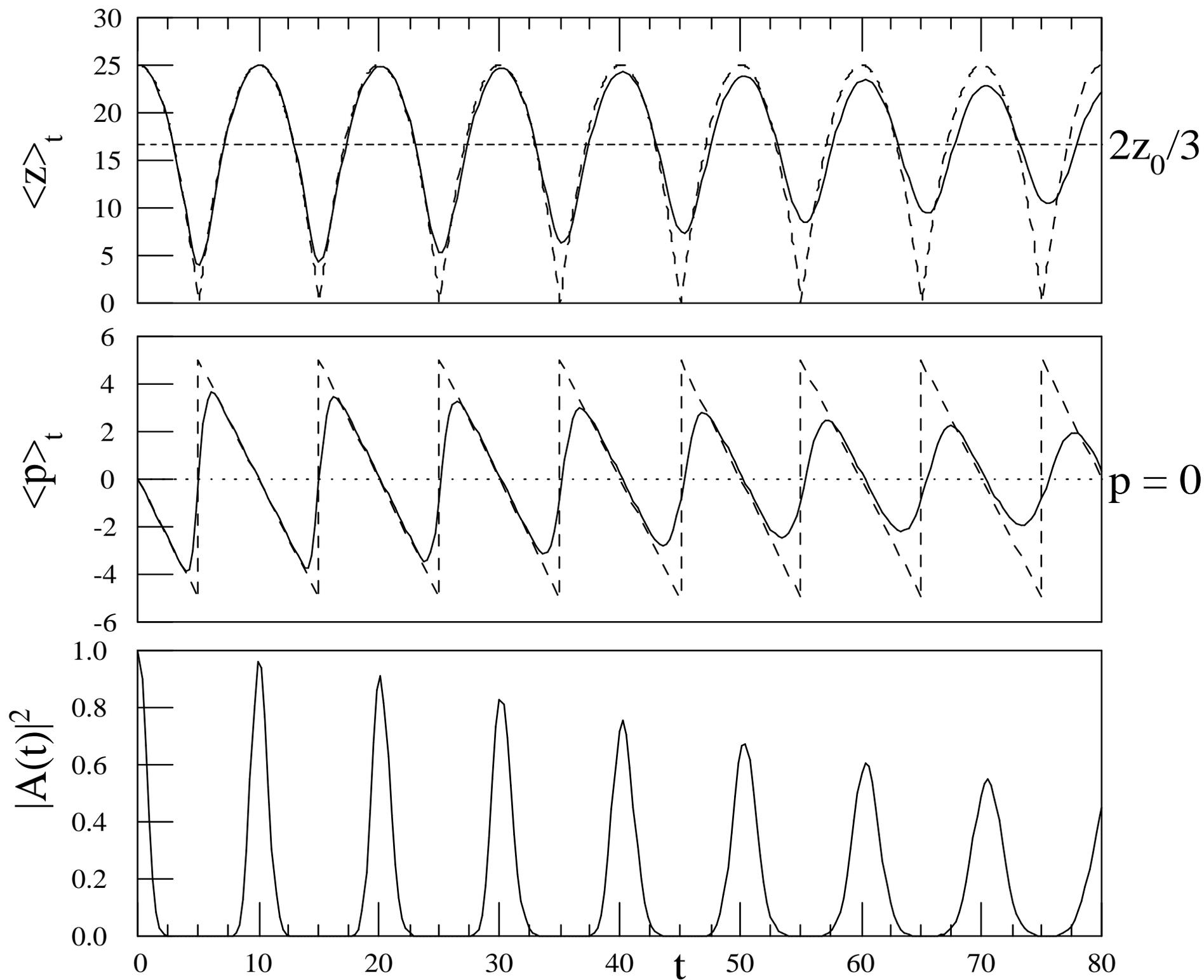
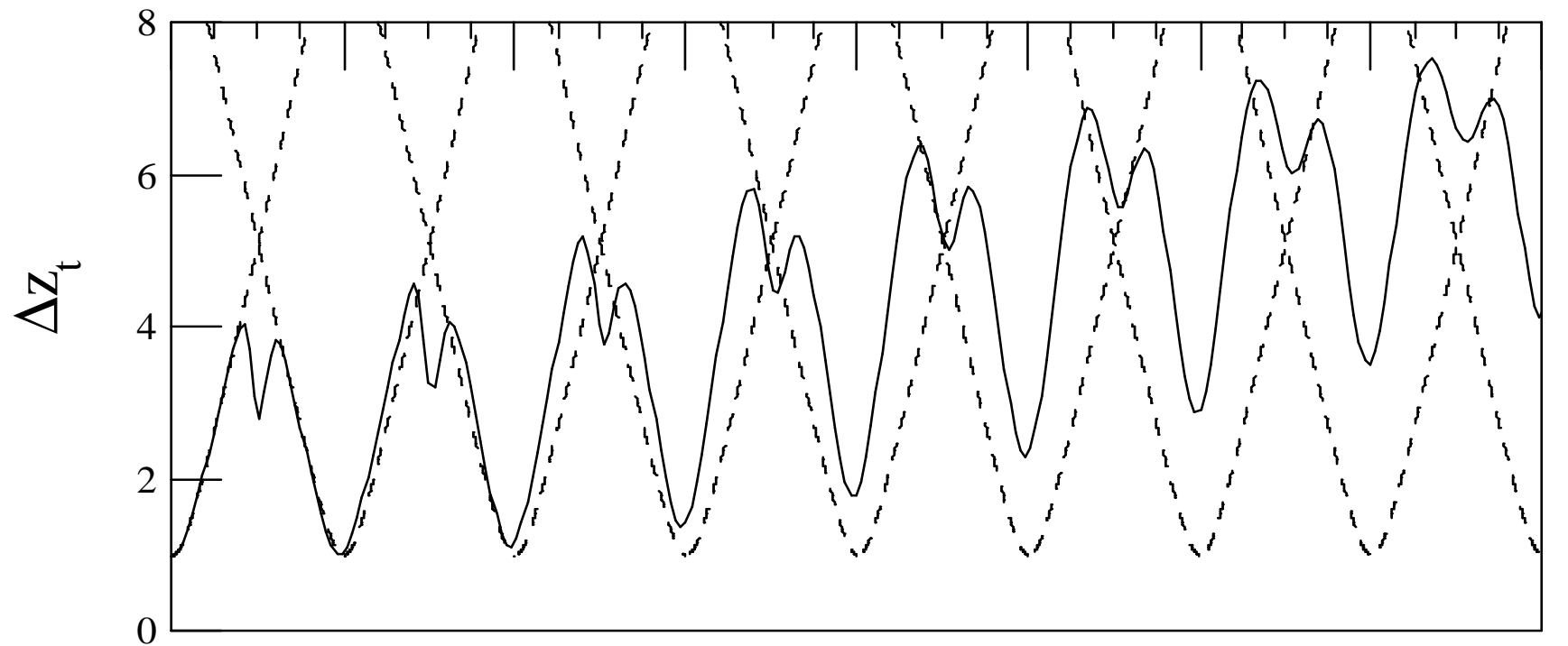
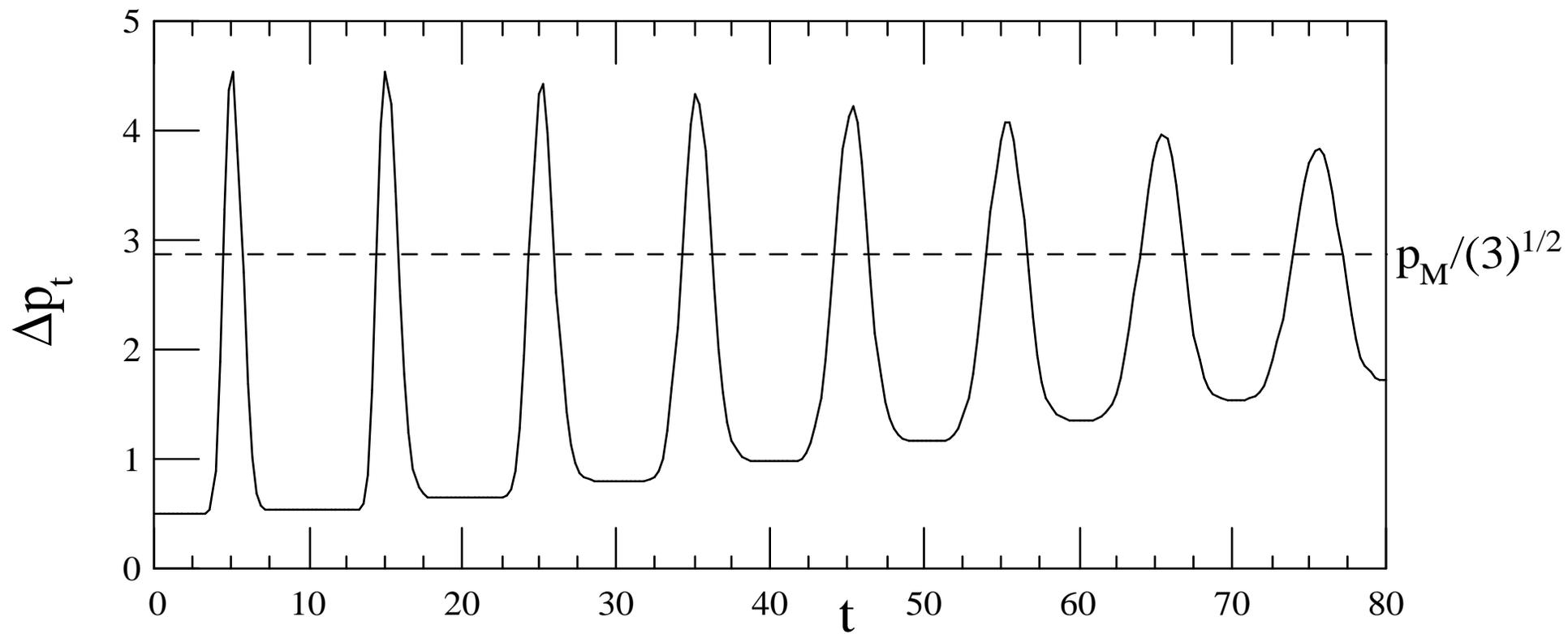
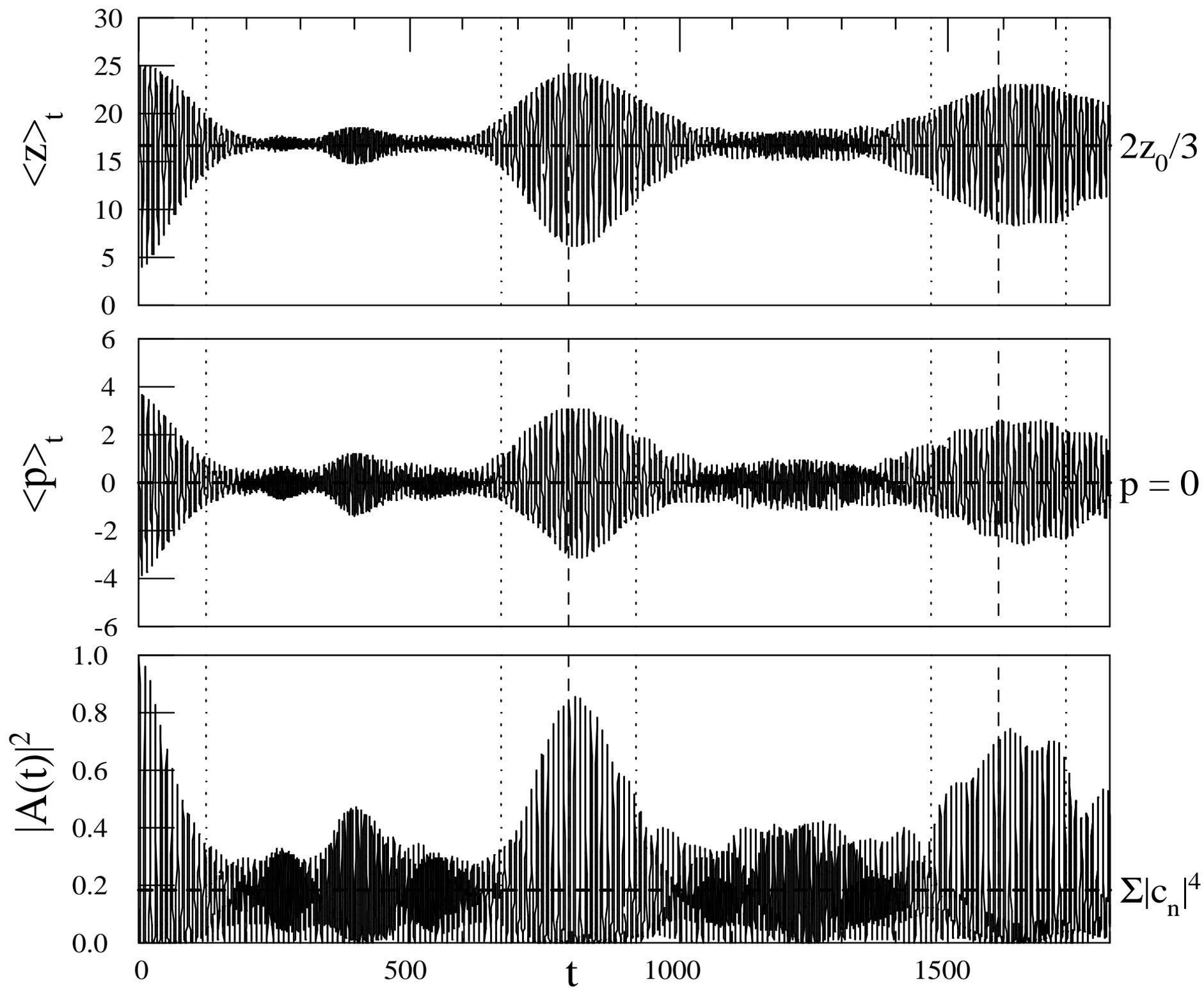
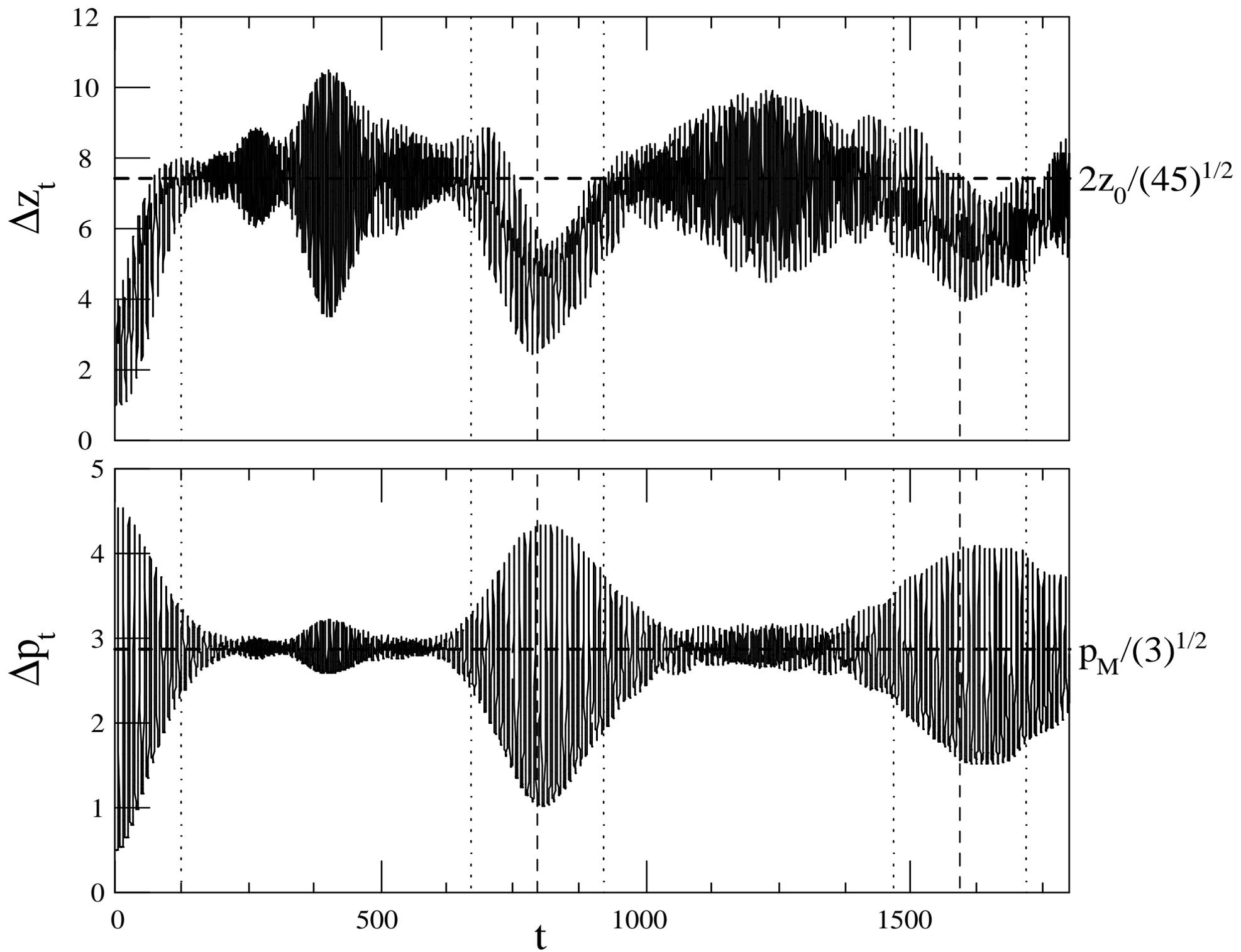
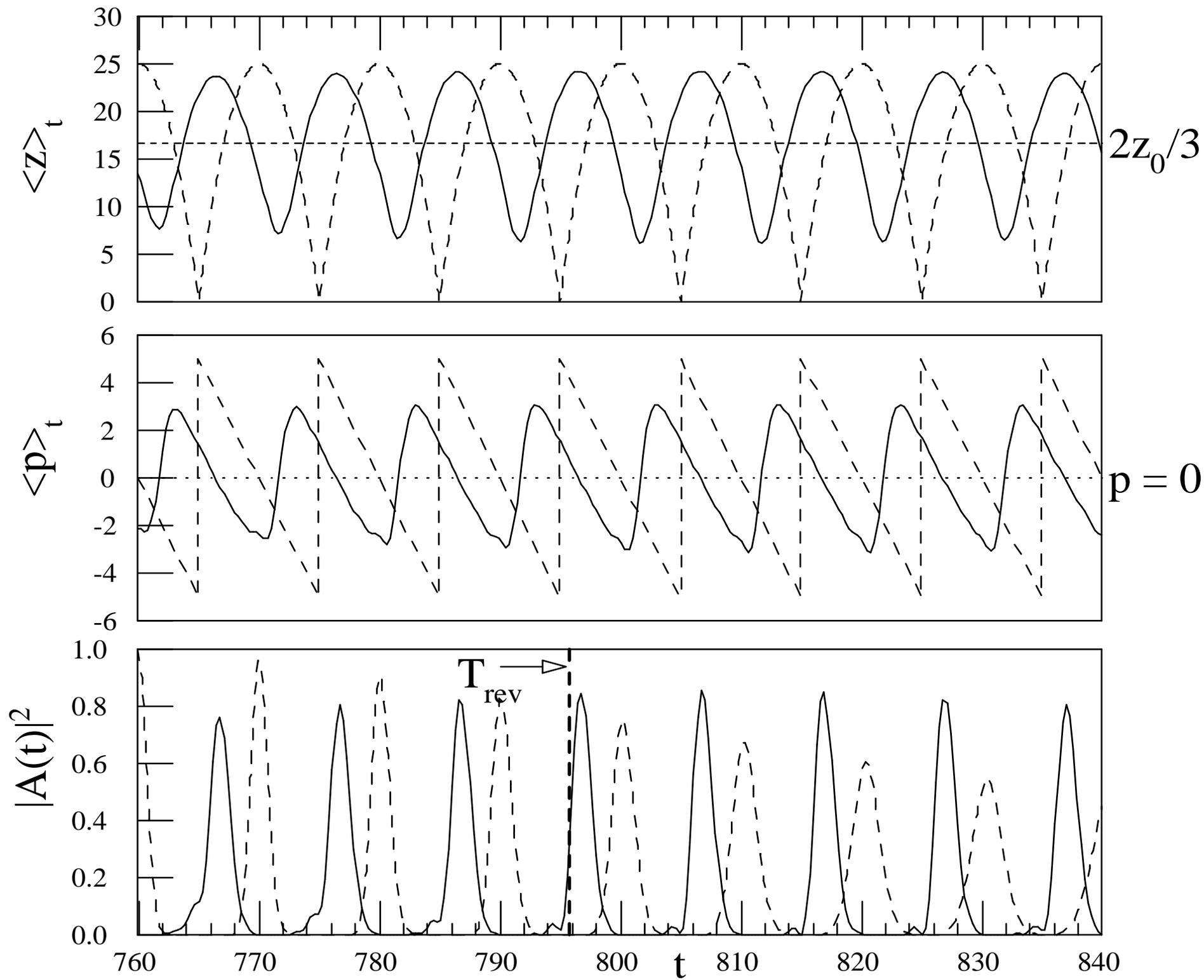
\end{document}